\documentclass[aps,longbibliography,notitlepage,nofootinbib,11pt, floatfix,tightenlines,%,superscriptaddress
]{revtex4-2}

\usepackage[utf8]{inputenc}
\usepackage[english,russian]{babel}
\usepackage{amsmath}
\usepackage{amssymb}
\usepackage{lipsum}
\usepackage{bm}
\usepackage{graphicx}
\usepackage{graphics}
\usepackage[dvipdf]{epsfig}
\usepackage{subcaption}
\usepackage[format=plain,labelfont={bf,small},textfont=small,justification=raggedright,singlelinecheck=false]{caption}
\usepackage{float}
\usepackage{wrapfig}
\usepackage{url}
\usepackage{epstopdf} 
\usepackage{xcolor}
\usepackage{tensor} % mixed notation etc
\usepackage{setspace}
\usepackage{commath} % norm
\newcommand{\bI}{{\mathbf I}}

\newcommand{\bV}{{\mathbf V}}

\newcommand{\bA}{{\mathbf A}}
\newcommand{\bB}{{\mathbf B}}
\newcommand{\bZ}{{\mathbf Z}}

\newcommand{\Tr}{{\textrm{Tr}}}
\newcommand{\Det}{{\textrm{Det}}}

\newcommand{\bb}{{\mathbf b}}
\newcommand{\bn}{{\mathbf n}}

\begin{document}

%\title{Corrigendum to, and remarks on, ``Dilation-invariant bending of elastic plates, and broken symmetry in shells'' and ``Energies for elastic plates and shells from quadratic-stretch elasticity'', Journal of Elasticity 153 571-579 and 581-598 }

\title{Assorted remarks on bending measures and energies for plates and shells, and their invariance properties}

\author{J. A. Hanna}
	\email{jhanna@unr.edu}
\affiliation{Department of Mechanical Engineering, University of Nevada,
    1664  N.\  Virginia  St.\ (0312),  Reno,  NV  89557-0312,  U.S.A.}
\author{E. Vitral}
	\email{vitralfr@rose-hulman.edu}
\affiliation{Department of Mechanical Engineering, Rose-Hulman Institute of Technology \\
  5500 Wabash Ave., Terre Haute, IN 47803, U.S.A. }

\begin{abstract}
In this note, we address several issues, including some raised in recent works and commentary, related to bending measures and energies for plates and shells, and certain of their invariance properties. 
We discuss overlaps and distinctions in results arising from two different definitions of stretching, correct an error and citation oversights in our prior work, % \cite{vitral2023dilation, vitral2023energies},
reiterate some of the early history of dilation-invariant bending measures, and provide additional brief observations regarding the relative size of energetic terms and the symmetrization of bending measures. 
A particular point of emphasis is the distinction between dilation-invariant measures and a recently introduced non-dilation-invariant measure for shells and curved rods. 
In the course of this discussion, we provide a simpler presentation of the elementary, but much neglected, fact that the through-thickness derivative of tangential stretch of material near the mid-surface of a thin body is the product of the mid-surface stretch and change in curvature.%, a quantity that  stands in contrast to measures that simply subtract the reference curvature from the product of stretch and curvature. 
\end{abstract}

% \pacs{}

\date{\today}

\selectlanguage{english}\maketitle
\selectlanguage{russian}

This note collects extended comments on aspects of our own and others' papers, particularly \cite{vitral2023dilation, vitral2023energies, Acharya00, Ghiba21, Virga24}.  
As such, we forego extensive introductory discussion, and expect the reader to be familiar with at least one of these works. 
We write this note as we believe certain issues regarding shell bending energies cannot be clearly understood from reading the aforementioned papers, because of their independent origins and lack of cross-awareness at the time of writing, and also because of a significant confluence of results arising from disparate motives and assumptions. %in the case of plates
%lack of cross-awareness 
%(our own papers overlooked earlier work by Acharya \cite{Acharya00} and recent work by B{\^{\i}}rsan and co-workers \cite{Birsan19}. )

Section \ref{measures} is concerned with an important distinction between two types of bending measures that may be obtained from different definitions of stretching or, for one type, from a dimensional reduction of a bulk energy. Section \ref{correction} corrects an error in our papers, which also provides us an opportunity to discuss, in both Sections \ref{correction} and \ref{symmetry}, other, minor points related to the size and nature of contributions from antisymmetric pieces of a bending energy.

\section{Remarks on different definitions of stretching, and resulting invariant bending measures}\label{measures}

In the first part of our work \cite{vitral2023dilation}, 
we introduced and employed a structure-dependent definition of stretching of thin bodies. 
The internal kinematics of such bodies is such that the normal (through-thickness) direction is constrained to behave differently than the tangential directions, so that such objects behave neither as unconstrained bulk three-dimensional continua nor as two-dimensional surfaces. 
We called ``pure stretching'' of such a body a deformation featuring through-thickness uniformity of stretch (ratio of deformed to referential length) of tangentially-oriented fibers.  By extension of this idea, an appropriate bending measure would be unchanged by such deformations.
This approach gave rise to certain bending measures and associated quadratic bending energies with a simple constitutive response, such that a pure moment is both linear in the bending measure and induces no stretching on a neutral surface. 
The location of this preserved neutral surface depends on the referential curvature, 
%cross sectional shape - we are really only considering rectangular-like shapes of a certain class here, from 'thickening' of a surface, 
but not the magnitude of the applied moment; for a flat plate or straight rod, it is the mid-surface, while for a curved rod it is displaced towards the center of referential curvature, in accordance with the classical result from linear elasticity. 
 % decoupled such that neutral surfaces were preserved under pure moments. 
%For example, a unidirectional moment applied to a flat plate or straight rod results in a mid-surface isometry.
%For a curved rod, the neutral surface lies on one side. 
An independent and equally important path to these same bending measures was taken in the second part of our work \cite{vitral2023energies}, in which we dimensionally reduced a 3D isotropic elastic energy quadratic in Biot-like strains, which are linear in the stretch tensor.  Regardless of the choice of energy, typical thin shell kinematics provides the form of the tangential stretch off of the mid-surface which leads to these bending measures, a point we will re-emphasize below.

In our work, we overlooked an earlier paper by Acharya \cite{Acharya00} in which the topic of bending measures was addressed by a different approach.  Acharya's approach employed the ``intuitive notion'' that a deformation does not involve bending if it preserves, up to a rigid rotation of the body, the field of normals to the mid-surface. 
Again, one can construct bending measures that are unchanged by such deformations.
 Recent work \cite{ghiba23, Virga24} indicates that a deformation of this type may be decomposed into a drilling rotation field that is everywhere oriented around the (referential) normals,\footnote{As our focus will be on the isotropic invariant content of certain bending measures, we will gloss over some aspects of tensorial measures. We refer the reader to the works \cite{Acharya00, Acharya24, Ghiba21, ghiba23, Virga24} for details, including discussions of the role of drilling rotations and comparisons between more and less stringent requirements on bending measures introduced by these other authors.}
 and a ``pure stretch'' defined such that the rotation field in the polar decomposition of the deformation gradient corresponds to rigid motion of the body.  %a simple example is a radial expansion, for which the rotation is the identity. %, so that it can be made to vanish in a certain frame.  
This definition of ``pure stretch'' is universal in that it does not distinguish between bodies that are bulk three-dimensional, purely two-dimensional, or thin structures with either plate- or shell-like reference curvature and associated internal kinematics. 
%the resulting definition of stretching/bending of a plate is the same as that for a curved shell. 
Bending energies quadratic in this other type of measure will also have a simple constitutive response, such that a pure moment is linear in the measure \cite{vitral2023energies}. 
It seems to have gone unremarked that such measures also give rise to a neutral surface; in this case, it is always the mid-surface, independently of referential curvature. 

Our structure-dependent definition of ``pure stretching'' of a particular surface does not correspond to a normal-preserving ``pure stretch'' of a general continuum, except in special coincident cases of objects with zero referential curvature, namely flat plates or straight rods. 
The use of the same term, ``stretching'', for both definitions is thus an unfortunate distraction from the underlying concepts. 
Although we do not know its origins, 
our impression is that Acharya's usage is established among many continuum mechanicians in the general context of bulk continua.  
%We were unaware of this at the time we initiated our work.  
Our coinage arose from the way the term ``stretching'' is informally used in studies of thin sheets in soft matter physics.  It could perhaps be called ``extension'' or something else.  That the symmetric positive definite part of the deformation gradient is called the stretch is, similarly, an accident of history. 

There are thus two different, but partially overlapping, requirements as to what is ``not bending'' and, associated with these, different types of bending measures unchanged by ``not bending'' deformations. 
One requirement is through-thickness uniformity of the stretch, in which the normals of a referentially-curved body will not be preserved 
%different elements along a referentially-curved body may experience different rotations 
 (the rotation will vary along the tangential directions, while through-thickness uniformity of rotation is part of the presumed kinematics of the thin body). 
% principal directions of stretch will vary alongside the tangents to a curved surface. 
The other requirement is preservation of the normals, in which the stretch will be nonuniform through the thickness of a referentially-curved body. 
These invariance requirements are not the same, but coincide for plates. 
Using \emph{either} requirement leads to the identification of dilations (also referred to as scaling) of \emph{plate} mid-surfaces as pure stretching. 
Note that the dilation is of the surface only--- the material on either side must follow kinematics appropriate to a thin body, and this is what dictates its stretch, as detailed in Figure 2 of \cite{vitral2023dilation}. 
In a dilation, these kinematics result in  %the stretch is isotropic, with no principal directions, but will lead to 
variations in stretch through a body with a curved reference shape. 
Thus, dilation of the mid-surface of a referentially-curved body is considered a combination of stretching and bending in our definition, but a pure stretching in the other definition. 
Conversely, a deformation that takes a body in a curved reference state and extends it at the same curvature, so that the stretch is uniform through the thickness, is considered a pure stretching in our definition, but a combination of stretching and bending in the other definition, as this process changes the normals attached to material points.

\subsection{Plates}

A simple dilation-invariant bending measure is the referential gradient of the (present) normal $\bar\nabla \bn$, a product of the deformation gradient and the present curvature, or a rotated product of the stretch and the present curvature. %a (rotated) product of the stretch and the present curvature.
In planar bending, we can use the scalar mid-surface stretch $\lambda$ and curvature $\kappa$ to write the corresponding dilation-invariant bending measure $\lambda\kappa$ or, equivalently, the material (referential arc length) derivative of the tangential angle.  This scalar measure and its invariance properties were introduced and recognized by Antman \cite{Antman68-2}.  %It corresponds to the referential gradient of the normal.  
To construct isotropic invariants to appear in bending energies, we may employ $\bar\nabla \bn$ or any other closely related tensor obtained by operations such as transposition, or rotation of one or more legs %or constituent tensors 
to/from the reference/present configuration; 
%all of which operations do not affect the scalar measures 
%operations that are irrelevant to the construction of the isotropic invariants that will appear in associated bending energies. 
%other than symmetrization these operations do not affect the scalar measures. 
for the present purposes, any of these encode the same physical information of interest. 
An early example of such a tensor was given by Atluri \cite[Equation 9.10]{atluri1984alternate}, who did not consider its invariance properties.  
Such invariance was recognized as important by Ghiba and co-workers \cite{Ghiba21}, and (a few months later) by the present authors, who considered a symmetrization of closely related referential and present tensors \cite{vitral2023dilation, vitral2023energies}. 
The concept of dilation invariance for plates, illustrated recently in Figure 6 of \cite{Ghiba21} and Figure 1 of \cite{vitral2023dilation}, was  introduced alongside Antman's original measure \cite{Antman68-2}--- see also the monograph by Antman \cite[Chapter 4]{Antman05} for further discussion of inflation of a ring and preservation of this bending measure within the context of a planar Cosserat rod.  
Other related tensors giving rise to the same isotropic invariants include the restriction to plates of one of the bending measures introduced by Acharya \cite[Equation 9]{Acharya00}, and the restriction to simple materials of the ``non-symmetric bending tensor'' of Neff \cite[last line of Equation 4.5]{Neff04}; %also last line of 5.4 and...? 
further history may be found in the references cited in \cite{vitral2023dilation, vitral2023energies}, noting however that this discussion missed \cite{Acharya00, Neff04}. 
Some of the same isotropic invariants will be obtained from the bending measure of Virga \cite{Virga24}, who introduces one of two possible symmetric ``squares'' of $\bar\nabla\bn$, namely the referential tensor $(\bar\nabla\bn)^\top\cdot\bar\nabla\bn$. 
The trace of either square is the continuum limit of the discrete energy found in \cite{SeungNelson88}; 
%While $(\bar\nabla\bn)^\top\cdot\bar\nabla\bn$ satisfies his requirements, $\bar\nabla\bn\cdot(\bar\nabla\bn)^\top$ does not, although they have the same invariants; their
 %, and satisfies the relaxed requirement of Ghiba and co-workers \cite{Ghiba21}. 
 %, while its determinant is higher order in stretch than quadratic. 
%Working with either squared quantity, one misses additional 
this will provide only some of the possible invariant content at quadratic order in stretch that can be obtained from tensors like $\bar\nabla \bn$.   %squaring the trace of an Atluri-type measure. 
For details of the other quadratic-stretch invariants, 
and those arising from reduction of a neo-Hookean energy or other quadratic-stretch energies, see Section 5.2.2 of \cite{vitral2023energies}.  
However, some other tensorial measures that provide the remaining quadratic-stretch scalar invariants do not satisfy certain of Virga's requirements related to drilling rotations of a \emph{deformed} plate \cite{Virga}. %, a point discussed in \cite{Ghiba21}? not clear, as drilling is referential for acharya, although only we seem to care about this distinction and what it means for material on either side of the midsurface.  
While Virga is concerned with purity of bending, our present concern is with purity of stretching. 
We note that some mid-surface stretching operations will be detected by both our and Virga's bending measures, and drilling operations (shear stretching) by our bending measures, but all only in the presence of a change in curvature, and this last is consistent with the idea that the bending content associated with changes in curvature may depend on tangential strains of a thin body. 
Extending or drilling at the reference curvature corresponds to through-thickness-uniform stretches, whereas extending or drilling at a different curvature does not, so the former will not contribute to our bending measure, but the latter will. 
In a body that has been deformed away from its reference curvature, particular additional combinations of extension and curvature change will not constitute bending in accordance with our definitions--- for a plate, this is dilation.  Presumably, there are also particular combinations of drilling and curvature change that do not constitute bending in accordance with our definitions.

\subsection{Shells}

In the case of curved shells or rods, the two requirements of through-thickness uniformity of stretch, or of preservation of normals, do not correspond. 
In the bending measures for shells that arise from one or the other of these requirements, the difference lies in what is being subtracted from the plate term.  The plate term will be something like the referential gradient of the normal $\bar\nabla\bn$, the product of stretch and present curvature, or some other closely related object.  
Our requirement leads us to subtract the product of stretch and referential curvature--- for details and discussion of our symmetrized referential and present tensors, %including the relations between various tensors mentioned above and below, 
see \cite[Sections 2-3]{vitral2023dilation} and \cite[Section 5.2]{vitral2023energies}. 
In planar bending, the corresponding scalar measure would be written 
$\lambda\left(\kappa - \bar\kappa\right)$.
For a curved reference state ($\bar\kappa \ne 0$), such measures are not dilation-invariant. 
They may be contrasted with dilation-invariant normal-preserving measures that subtract the referential curvature (or a closely related tensor); in planar bending, the corresponding scalar measure would be
 $\lambda\kappa - \bar\kappa$, which also appears in Antman \cite{Antman68-2}. 
An asymmetric tensorial measure giving rise to the former non-dilation-invariant $\lambda\left(\kappa - \bar\kappa\right)$ type of scalar invariants, when restricted to simple materials, appeared a couple of years earlier than our work in a paper by B{\^{\i}}rsan and co-workers that we overlooked; for initial implicit appearance in an energy, see \cite[Equation 98 or 104]{Birsan19}, and for explicit display see \cite[Equation 126]{Birsan20} % (Birsan's email cites eqs 53,68,126,and section 5.3)
 or the ``new mixed bending tensor'' in \cite[Equations 122-123]{Birsan21}. 
 This tensor, re-expressed using an additional ``fictitious planar configuration'', also appears in \cite[Equation 5.4]{Ghiba20}, but these authors do not dwell on its significance, choosing instead to rewrite it as  
%however they didn't recognize how important their new tensor was, instead for whatever reason 
a different tensor minus a tensor of %Antman/Atluri
 the latter dilation-invariant $\lambda\kappa - \bar\kappa$ type, whose properties they proceed to explore and discuss at length in  \cite{Ghiba21}. 
 This latter tensor, as seen in \cite[last line of Equation 5.5]{Ghiba20} or \cite[Equation 5.9 or 6.2]{Ghiba21}, is renamed the ``change of curvature'' tensor in \cite{ghiba23}, although it (\emph{via} the aforementioned fictitious configuration) effectively subtracts referential curvature from the product of stretch and curvature, not from the curvature. 
This splitting of the original tensor is akin to simply replacing the offending non-dilation-invariant subtracted piece by a dilation-invariant one, that is, rewriting $\lambda\left(\kappa - \bar\kappa\right)$ as $-\left(\lambda-1\right)\bar\kappa +\left(\lambda\kappa - \bar\kappa\right)$ and proceeding to focus only on the rightmost parenthetical term.  
%they proceed in Ghiba21 to discuss R (the term on the right hand side) and its properties, but the energy is quadratic in the non-dilation-invariant quantity that is akin to ours.  
Tensors giving rise to the latter dilation-invariant $\lambda\kappa - \bar\kappa$ type of scalar invariants appear in several places.  An early example was also given by Atluri \cite[Equation 9.25]{atluri1984alternate}, implicitly in a two-dimensional Biot-like energy.  
Other closely related tensors include the aforementioned measure of Acharya \cite[Equation 9]{Acharya00}; further history may be found in the references cited in \cite{vitral2023dilation, vitral2023energies}, noting however that this discussion missed \cite{Acharya00, Ghiba20}. 

%related to refgradn minus refk (referential gradient of the normal, or the product of stretch and curvature, minus the referential curvature.) by by transposition, or rotation of one or more legs %or constituent tensors 
%to/from the reference/present configuration (or some fictitious flat configuration), operations that are irrelevant to the construction of the isotropic invariants that will appear in associated bending energies. 

A non-dilation-invariant tensor encoding the product of the mid-surface stretch and the change in mid-surface curvature also arises naturally as the additional (tangential) stretch of material located just off of the mid-surface. 
The manner in which this enters a reduction of a 3D quadratic-Biot-type bulk energy, \emph{via} the through-thickness derivative of the deformation gradient, can be seen 
%from the expansion around the mid-surface of the tangent vectors comprising the deformation gradient in convected coordinates,
 in Equations 27-31 of \cite{vitral2023energies}, noting that the coefficient $\alpha_1$ appearing in these expressions is unity at lowest order in stretch. 
 This is also the context in which B{\^{\i}}rsan and co-workers arrive at their tensor, whose appearance as a through-thickness strain derivative, and subsequently as a squared quantity in a reduced energy, is visible in \cite{Birsan19, Birsan20, Birsan21}; they consider a more complex material and kinematics with additional degrees of freedom, which can be restricted to a simple quadratic-Biot-type material. 
 %Their approach starts with a more complex material and kinematics, employs different assumptions, and follows different procedures 
%We note that that they have not yet performed the reduction directly from a 3D Biot-type energy to a 2D Biot-type energy.   but seems that ``diagram is commuting''. 
%It is curious that, 
While a planar or axisymmetric version of this type of derivation was performed in several places for flat bodies \cite{IrschikGerstmayr09, OshriDiamant17, WoodHanna19}, 
we are not aware of such a simple derivation for curved rods in the literature.
%(perhaps because very few work with Biot-like strains, see Neff08 and VitralHanna22 for recent examples.)
Because of this apparent lacuna, we provide a sketch of this calculation below.

The natural appearance of non-dilation-invariant $\lambda\left(\kappa - \bar\kappa\right)$ type of measures can be understood from the following informal and incomplete planar derivation in the style of \cite{IrschikGerstmayr09, WoodHanna19}, or implicitly through computation of through-thickness-uniform stretches in \cite{vitral2023dilation}, or explicitly and in full tensorial form in \cite{vitral2023energies}--- a detailed and more careful approach built on \cite{OzendaVirga21, steigmann2013koiter} that leads to the same results--- or alternatively through the thorough derivations of B{\^{\i}}rsan and co-workers \cite{Birsan19, Birsan20, Birsan21}.

Consider a thin body in the plane, with uniform thickness $h$, parameterized by material coordinates $x$ and $z$ in the tangential and normal directions, respectively.  
Let $x$ be the arc length of the referential mid-line. The deformed mid-line has stretch $\lambda$, tangential angle $\theta$, and scalar curvature $\kappa = d_x\theta/\lambda$.  The corresponding referential quantities are $\bar\theta$ and $\bar\kappa = d_x \bar\theta$. The angular description is not necessary, but included for completeness.

The tangential stretch is the ratio of lengths of deformed and referential line elements in the $x$ direction.  
Considering shear-free deformations for simplicity, this stretch varies with distance from the mid-surface approximately as 
\begin{align}
	\frac{l (z)}{\bar l (z)} \approx \frac{l (0) \left[ 1-zd_x\theta/\lambda \right] }{   \bar l (0) \left[ 1-zd_x \bar\theta \right]}  =
	\lambda \, \frac{1-z\kappa}{1-z\bar\kappa} 
	\approx \lambda \left[1 - z \left(\kappa-\bar\kappa\right)\right] \, . \label{stretch}
\end{align}
The first approximation corresponds to neglect of the Poisson effect in the $z$ direction, which forms part of the classical Kirchhoff-Love kinematic assumptions and remains in their more consistent generalization \cite{OzendaVirga21}, while the second approximation corresponds to small $z\kappa$ and $z\bar\kappa$. 
%\com{draw figure?} 
Thus, we already see that, to a first approximation, the normal (through-thickness) derivative of the tangential stretch is 
\emph{the product of the mid-surface stretch and change in curvature}.  This quantity will then appear in strains and energies of any order of stretch. 
Further kinematic assumptions allow the transverse stretch to be related to the tangential stretch, so the energy can be written in terms of the latter only. 
Merely as an example, we may consider an energy quadratic in Biot strains, which are obtained by subtracting unity from \eqref{stretch}; on the mid-line we define $\Delta = \lambda - 1$. 
This energy would be of the form
\begin{align}
	\mathcal{E} \sim \int\! \textrm{d}x \int_{-h/2}^{h/2} \textrm{d}z \left(1-z\bar\kappa\right)\left[ \Delta - z\lambda\left(\kappa-\bar\kappa\right) \right]^2 \, , 
\end{align}
whose leading-order terms for small strain and thickness are of the form $h\Delta^2$ and $h^3\lambda^2\left(\kappa-\bar\kappa\right)^2$. 
%The additional question of whether a symmetric or asymmetric tensorial measure might arise in a general derivation is addressed later in Section \ref{symmetry}.  
%\com{is neohookean the same bending term? not stretching}

Perhaps if such a simple calculation had been performed, the bending measure $\lambda\left(\kappa-\bar\kappa\right)$ might have appeared and been adopted earlier, independently of any considerations of through-thickness stretch uniformity. 
Before we had performed either the stretch uniformity \cite{vitral2023dilation} or reduction \cite{vitral2023energies} calculations, we were also ready to extend the property of dilation invariance to shells, misguided by its elegance, instead of allowing the geometry of the shell to break this symmetry.  
Considering internal kinematics appropriate to a thin body, rather than proposing a direct theory of surfaces, leads to quantities that distinguish between bending a plate that has already been bent into a curve, and bending a curved shell at rest. 
%We note, however, that such internal kinematic considerations, appropriate to a thin body, would not be expected in the development of direct theories of two-dimensional surfaces. 
In contrast, dilation-invariant curved rod or shell measures have been proposed in various places, but never derived through a reduction.  They have been introduced either as a direct prescription, or obtained as the result of imposing dilation invariance or normal preservation from the beginning. 
%assumed based on symmetry or beauty or whatever -- starting with certain requirements already begs the question of the measure. dilation invariance from beauty. normal preservation from carryover of bulk definitions. 
Such requirements do not distinguish between bent plates and shells at rest. 

However, as already mentioned above, either type of measure satisfies our original goals in \cite{vitral2023dilation, vitral2023energies} of obtaining a moment linear in the measure, and a neutral surface.  In the non-dilation-invariant case, the measure is also obtainable from a reduction, and the neutral surface is consistent with classical results for a curved beam.  In the dilation-invariant case, the neutral surface is the mid-surface, which is perhaps more convenient when constructing a shell theory. 
The two types of measures agree for flat plates and straight rods.

\section{Correction, with a comment on the size of relevant terms}\label{correction}

In the papers \cite{vitral2023dilation,vitral2023energies}, we discuss a plate bending measure $\text{\Large{\bf{\cyrl}}}= \textrm{sym}(\bV\cdot\bb)$ formed by symmetrizing the asymmetric product of two symmetric tensors, the left stretch $\bV$ and curvature $\bb$. 
In both papers, a similar incorrect statement is made in the discussion of the results, to the effect that the determinant of $\text{\Large{\bf{\cyrl}}}$ is the product of determinants of $\bV$ and $\bb$, which are respectively the Jacobian determinant $J$ and Gau{\ss}ian curvature $K$. 
This elementary mistake in linear algebra was never employed in any of the derivations, but contributed to incorrect conclusions about the geometric nature of certain terms in a bending energy, as expressed in two places in our texts. 

Let $\bA$ and $\bB$ be second-order tensors. While it is true that $\Det(\bA\cdot\bB)=\Det(\bB\cdot\bA) = \Det\,\bA\,\Det\,\bB$, this is not equal to $\Det( \textrm{sym}(\bA\cdot\bB) )$ %= \Det ( \tfrac{1}{2}(\bA\cdot\bB + \bB\cdot\bA) )$ 
unless $\bA$ and $\bB$ commute.
For tensors of dimension two, the determinant of the sum and difference of tensors $\bA$ and $\bB$ can be expressed using (see \cite[Section 9]{truesdell2004nonlinear} for several more general relevant expressions): 
\begin{align}
\Det(\bA\pm\bB) &= \Det\,\bA\, \Det( \bI \pm\bA^{-1}\cdot\bB )   \nonumber \\
	&= \Det\,\bA (1 \pm \Tr(\bA^{-1}\cdot\bB) + \Det(\bA^{-1}\cdot\bB) ) \nonumber \\
	&= \Det\,\bA \pm \Tr( \Det(\bA) \bA^{-1}\cdot\bB) + \Det\,\bB \, .
\end{align}
Alternately, we may use the relations \cite{Virga}
\begin{align}
2\Det(\bA\pm\bB) &= (\Tr(\bA\pm\bB))^2 - \Tr((\bA\pm\bB)\cdot(\bA\pm\bB)) \nonumber \\
	&= (\Tr\,\bA)^2 \pm 2\Tr\,\bA\,\Tr\,\bB + (\Tr\,\bB)^2 - \Tr(\bA\cdot\bA) \mp 2\Tr(\bA\cdot\bB) - \Tr(\bB\cdot\bB) \nonumber \\ 
	&= 2\Det\,\bA + 2\Det\,\bB \pm 2\Tr\,\bA\,\Tr\,\bB \mp 2\Tr(\bA\cdot\bB) \, .\label{virga}
\end{align}
Combining either pair of expressions, we find that 
\begin{align}
\Det(\bA+\bB) + \Det(\bA-\bB)  =2(\Det\,\bA+\Det\,\bB)\,, 
 \end{align}
and, therefore, $\Det$ commutes with the decomposition into symmetric and antisymmetric parts, 
\begin{align}
\Det( \textrm{sym}(\bA\cdot\bB) ) + \Det( \textrm{skew}(\bA\cdot\bB) )  =
2( \Det\tfrac{1}{2}(\bA\cdot\bB)+ \Det\tfrac{1}{2}(\bB\cdot\bA) ) = \Det(\bA\cdot\bB) = 
 \Det\,\bA\,\Det\,\bB\,. \label{detsym}
 \end{align}
 For the particular case at hand, 
%note $\bV$ and curvature $\bb$ are symmetric, but the Atluri measure $\bV\cdot\bb$ is not
employing the definition \cite{VitralHanna22} of the Bell strain $\mathbf{E}_{\textrm{Bell}} = \bV - \mathbf{I}$, we have 
\begin{align}
  \Det\,\text{\Large{\bf{\cyrl}}}  &= \Det\,\bV\,\Det\,\bb - \Det ( \textrm{skew}(\bV\cdot\bb) )  \nonumber \\
  &= JK - \Det ( \textrm{skew}(\mathbf{E}_{\textrm{Bell}}\cdot\bb) )   \, .
%  &= J\,K - \Det\tfrac{1}{2}(\mathbf{E}_{\textrm{Bell}}\cdot\bb-\bb\cdot\mathbf{E}_{\textrm{Bell}})
 \label{eq:detlu}
\end{align}
The second term only vanishes if the principal axes of stretch and curvature align (if these two symmetric tensors commute, they are coaxial).  

The relevant errors are those referring to the determinant of the plate bending measure $\text{\Large{\bf{\cyrl}}}$ and its referential-area-weighted integral: in \cite{vitral2023dilation}, Section 2, the paragraph following the paragraph containing Equation 4, and
%\textcolor{blue}{It is interesting to note that the referential-weighted determinant of \textcolor{red}{(3)} is a purely geometric quantity, the integral of Gaussian curvature $\int dA \,\textcolor{red}{i_2} = \int dA\,\textcolor{red}{\Det \,\mathbf{V} \,\Det \,\bb} = \int da \,\Det \,\bb$, which is equal to a boundary term plus a topological invariant, and therefore constant for any closed surface. This is curious because, again, we don’t expect an elastic energy to be geometric. It further implies that SN or neo-Hookean models may sufficiently capture the behavior of closed surfaces composed of quadratic-stretch plate-like material, however it is more likely that a study of a closed surface will be concerned with a shell energy instead of a plate energy.}
in \cite{vitral2023energies}, Section 5.2.1, the final paragraph.
%\textcolor{blue}{Under isometric deformations of the mid-surface, we have $\text{\Large{\bf{\cyrl}}}=\bb$ and thus the geometric quantities $\Tr\,\text{\Large{\bf{\cyrl}}} = 2H$ and $\Det\,\text{\Large{\bf{\cyrl}}} = K = 0$ will appear in the energy density obtained from the special case of (44) for isometric plates. In general, $\textcolor{red}{\Det\text{\Large{\bf{\cyrl}}} = \Det\, \mathbf{V}\, \Det \,\bb = JK}$ where the Jacobian determinant $J = \sqrt{a/A}$ and $da = \sqrt{a}\, d\eta^1d\eta^2$ is the present area form. Thus, part of the plate bending energy is a purely geometric quantity, $\int dA \,\textcolor{red}{\Det\text{\Large{\bf{\cyrl}}}} = \int dA\,\textcolor{red}{JK} = \int da\, K$ , which is equal to a boundary term plus a topological invariant.}

%\section{Comment on the size of determinant terms}

The determinant of any symmetric plate bending measure linear in curvature will contain $K$, which is related to the derivatives of the metric.  
%note that TE relates derivatives of the metric with K.  
The referential $\bar K$ of a plate is zero. 
Changes in metric, and thus $K$, scale as stretch squared minus one, and are thus linear in $\Delta$, where $\Delta$ is an eigenvalue of the Bell strain (or equivalently the more common Biot strain that employs the right stretch).
The first term in \eqref{eq:detlu} is $O(\Delta)$ while the second is $O(\Delta^2)$. 
 This determinant of the antisymmetric piece can be compared with the terms coming from the square of the trace of the symmetric bending measure, which will be $O(1)$ with $O(\Delta)$ corrections.
These terms can be considered in the context of an %areal
 energy density that is an expansion in small thickness $h$ and small strains $\Delta$, with a stretching term of $O(h\Delta^2)$ and a bending term of $O(h^3B^2)$ (where $B$ is some suitable eigenvalue of the bending measure), in which we are already dropping terms coming from Poisson coupling to bending of $O(h^3\Delta B^2)$, among terms of various other orders. 
 %approximation with one parameter models?
 In shells, which provide additional complications, we also often ignore terms coming from the expansion of the area form around the mid-surface, of similar order $O(h^3\Delta \bar H B)$ (where $\bar H$ is the referential mean curvature). 

All of this is to say that arguments might be made to ignore some terms for small strain theories, should this be desirable for other reasons.  It is also worth remembering that the distinctions being made between bending measures are small in magnitude for small-strain theories (comparable for example to the neglected Poisson terms), their importance being primarily in keeping certain natural objects intact, and in producing a certain qualitative response.

 \section{Remarks on antisymmetric contributions to the energy}\label{symmetry} 
 
In our work \cite{vitral2023dilation,vitral2023energies}, we construct the most general isotropic quadratic energy for symmetric tensors such as the plate measure $\text{\Large{\bf{\cyrl}}}= \textrm{sym}(\bV\cdot\bb)$, which includes  
 $(\Tr\,\text{\Large{\bf{\cyrl}}})^2$ and $\Tr(\text{\Large{\bf{\cyrl}}}\cdot\text{\Large{\bf{\cyrl}}})$, or equivalently 
$(\Tr\,\text{\Large{\bf{\cyrl}}})^2$ and $\Det\,\text{\Large{\bf{\cyrl}}}$. 
 We also derive such an energy by reduction from a symmetric 3D strain \cite{vitral2023energies}.  However, the latter process requires care to preserve the symmetry of the measure. 
 The symmetric stretch can be expressed in two complementary ways, either using the rotation and deformation gradient, or their transposes. We take derivatives of this quantity through the thickness of the body, and neglect as higher-order the derivative of the rotation (see Equation 5 and the text in the short paragraph spanning pages 587-588; to see how this strain measure enters the energy, the relevant Equations are 18-20, 23-24). This process can potentially result in the spurious introduction of asymmetry into this quantity, unless it is expressed symmetrically, as is done in Equation 31 of \cite{vitral2023energies}. 
  
  If we, instead, directly begin with an asymmetric tensor, the most general isotropic quadratic energy has three parameters (here $\lambda$, $\mu$, $\alpha$), as the tensor can be ``squared'' with itself or with its transpose. 
The asymmetric tensor $\bV\cdot\bb$, the rotated referential gradient of the normal that appears in some manner in all of the plate bending tensors discussed in this note, has some additional structure, as it is formed from two symmetric tensors $\bV$ and $\bb$ (so that $(\bV\cdot\bb)^\top = \bb\cdot\bV$).  
In this case, the most general form of an isotropic quadratic energy is, after arranging to isolate the antisymmetric piece, and further rewriting compactly using $\text{\Large{\bf{\cyrl}}}= \textrm{sym}(\bV\cdot\bb)$ and its counterpart $\bZ= \textrm{skew}(\bV\cdot\bb)$, 
\begin{align}
	&\lambda (\Tr(\bV\cdot\bb))^2  + \mu ( \Tr(\bV\cdot\bb\cdot\bb\cdot\bV) + \Tr(\bV\cdot\bb\cdot\bV\cdot\bb) ) + \alpha ( \Tr(\bV\cdot\bb\cdot\bb\cdot\bV) - \Tr(\bV\cdot\bb\cdot\bV\cdot\bb) ) \nonumber  \\
	&= \lambda (\Tr\,\text{\Large{\bf{\cyrl}}})^2 + 2\mu\Tr(\text{\Large{\bf{\cyrl}}}\cdot\text{\Large{\bf{\cyrl}}}) - 2\alpha\Tr (\bZ\cdot\bZ) \, , \label{threeterms}
\end{align}
 noting further the two-dimensional relations 
 %It can be seen that the symmetric $\mu$ term is
 $\Tr(\text{\Large{\bf{\cyrl}}}\cdot\text{\Large{\bf{\cyrl}}}) = (\Tr\,\text{\Large{\bf{\cyrl}}})^2 - 2\Det\,\text{\Large{\bf{\cyrl}}}$ and
 %and the antisymmetric $\alpha$ term is 
 $-\Tr (\bZ\cdot\bZ) = \Tr (\bZ\cdot\bZ^\top) = 2\Det\,\bZ$. 
We remark that the original tensor $\bV\cdot\bb$ has two invariants, and it is possible to re-express the three terms in \eqref{threeterms} in terms of these using \eqref{detsym}, that is, $\Det\,\text{\Large{\bf{\cyrl}}} + \Det\,\bZ = \Det(\bV\cdot\bb)$.

What information is contained in the antisymmetric term?   
The determinant only provides a small, positive term of $O(\Delta^2)$; expressed in terms of Cartesian components it is \newline $( (V_{11}-V_{22})b_{12} - (b_{11}-b_{22})V_{12} )^2$.  %(looks like $\alpha$ must be nonnegative)
This couples the deviatoric parts (difference of eigenvalues) of the stretch and curvature. 
Consider equal stretching and compression of a square element, resulting in pure shear along a diagonal, followed by bending into a saddle with equal-magnitude principal curvatures along the diagonals.  The principal bending is along the unstretched (sheared) directions, and the principal stretching is along the unbent directions of the saddle.
A positive $\alpha$ corresponds to a positive ``interaction energy'' between the two deformations. %, which seems to suggest some kind of tension-compression or up-down-bending asymmetry... are there physical grounds on which to set this to zero?
%should there be any additional energy associated with bending in an unstretched direction, or stretching in an unbent direction? 
We reserve consideration of the physical meaning, origins, and permissibility of such terms, and their counterparts in the more complex case of shells, for another time.

%\com{can we comment further on why Virga doesn't like the trace of atluri?} 

\section{Conclusions}
We have compared two types of shell bending energy with simple constitutive response, corresponding to different definitions of stretching and different invariance properties, which coincide for plates.  
This included a sketch of a derivation indicating how a dimensional reduction gives rise to one of these energies. 
We have also corrected an error and citation oversights in our own work, and briefly discussed energetic terms 
related to asymmetries in bending measures.

\section*{Acknowledgments}
We are grateful to E. G. Virga for alerting us to the error corrected in Section \ref{correction}, and for helpful input on this note, including suggesting the simple derivation \eqref{virga}. 
We also thank M. B{\^{\i}}rsan and P. Neff for helping us understand their results. 
We acknowledge U.S. National Science Foundation grant CMMI-2001262. 

%\appendix
%\selectlanguage{english}

%\bibliographystyle{unsrt}
%\bibliography{refs_remarks}

\end{document}